# Sensitivity of Microwave Interferometer in the Limiter Shadow to filaments in ASDEX Upgrade

Mariia Usoltceva[1], Stéphane Heuraux[2], Ildar Khabibullin[3], Helmut Faugel[1], Helmut Fünfgelder[1], Vladimir Bobkov[1] and ASDEX Upgrade Team[1]

[1] Max-Planck-Institut für Plasmaphysik, Boltzmannstr. 2, 85748 Garching, Germany
[2] Université de Lorraine, CNRS Institut Jean Lamour BP50840, F-54011 Nancy, France
[3] Max-Planck-Institut für Astrophysik, Karl-Schwarzschild-Straße 1, 85748 Garching, Germany

E-mail: mariia.usoltceva@ipp.mpg.de

**Abstract**

Microwave interferometer in the Limiter Shadow (MILS) is a new diagnostic, installed on ASDEX Upgrade for electron density measurements in the far Scrape-Off Layer (SOL). At the chosen frequency of 47 GHz the region of measurements varies within several centimeters before and after the limiter, depending on the density. 200 kHz data acquisition allows resolving transient events such as edge localised modes (ELMs) filaments and turbulence filaments. The measured quantities, phase shift and power decay of the microwave beam, which crosses the plasma, are directly connected to the density and do not depend on any other plasma quantity. In this work, we analyse the influence of a filamentary perturbation on MILS signals. Simple representation of a filament is adopted, with parameters relevant to experimental filament properties, reported for ASDEX Upgrade. Forward modelling is done in COMSOL software by using RAPLICASOL, to study the response of the MILS synthetic diagnostic to the presence of a filament. Qualitative and quantitative dependencies are obtained and the boundaries of MILS sensitivity to filaments, or to the density perturbation in far SOL in general, are outlined.

## 1. Introduction

In the Scrape-Off-Layer (SOL) of a tokamak, the particle transport and the resulting density distribution are largely influenced (or often defined) by filaments. Such intermittent transport events can be large when assosiated with an edge localised mode (ELM) or small when they originate from the turbulence between ELMs or in an L-mode. Properties of filamentary transport have been a subject of an extensive research [1], [2].

Microwave interferometer in the Limiter Shadow (MILS) [3] is a new diagnostic, installed on ASDEX Upgrade for electron density measurements in the far Scrape-Off Layer. At the chosen frequency of 47 GHz the region of measurements lies within several centimeters before and after the limiter. 200 kHz data acquisition allows resolving important transient plasma edge events. The measured quantities, phase shift and power decay of the microwave beam, which crosses the plasma, are directly connected to the density and, for the well-aligned O-mode configuration, do not depend on any other plasma quantity.

Forward modelling is used for the reconstruction of the plasma density profile in far SOL from measured signals [3]. This method is applicable to a monotonic inter-ELM like density profile without large perturbations. In this work, we aim at studying the influence of the filament presence on the signals of MILS. Such description should be closer to experimental conditions, where turbulence and blobs are always present.

A simplified representation of the filaments is adopted in the current study, where they are defined as circular or bean-shaped structures, elongated along the toroidal field lines. By simulating filaments with various sizes and density levels, we explore the range of the detection of such structures and find correlations between the filament properties and the MILS signals of the synthetic diagnostic.

**2. Experimental and synthetic diagnostics**

MILS consists of two microwave horn antennas, emitter and receiver, aligned and directed at each other, with 42.5 cm distance between them (Fig. 1a). The axis of the interferometer is oriented perpendicularly to the background magnetic field and the wave at $f$ = 47 GHz is sent in the O-mode polarization, making the measured signal sensitive to the density only.

In the synthetic diagnostic [3], the wave propagation is simulated in a limited region between the horns, where it matters for the received signal. The full-wave 3D modelling is done in COMSOL by using RAPLICASOL approach of wave propagation simulation in cold plasma [4], where absorbing material surrounds the plasma domain and ensures a single-pass propagation. Fig. 1b shows the radial-poloidal cross-section of the simulation domain with an example of wave propagation in plasma with density exponentially increasing towards the core, as described in details in [3]. The core plasma is on the left, with the flux surfaces approximated by circles concentric to the circle that defines the limiter.

In the toroidal direction, the horn antennas of MILS have the size of 2 cm. The signal is therefore collected from a very thin toroidal region of the same size. On such scale we can assume the variation of plasma density in toroidal direction as negligible.

The radial span of the density profile, which has influence on the collected signal, is much larger, due to the fact that the part of the beam, which goes towars the high density plasma, gets refracted and is able to reach the receiver (Fig. 1b). The signal is therefore collected from a large volume, it experiences interference inside the receiver horn and converts into the waveguide fundamental mode, which propagates to the detecting hardware. The previous study of the measurement region of MILS shows that the part of the radial density profile, which influences the MILS measurements, has a length of ~ 5 cm and is located near the limiter edge, more towards the core plasma for lower densities and towards outside for higher densities [3].

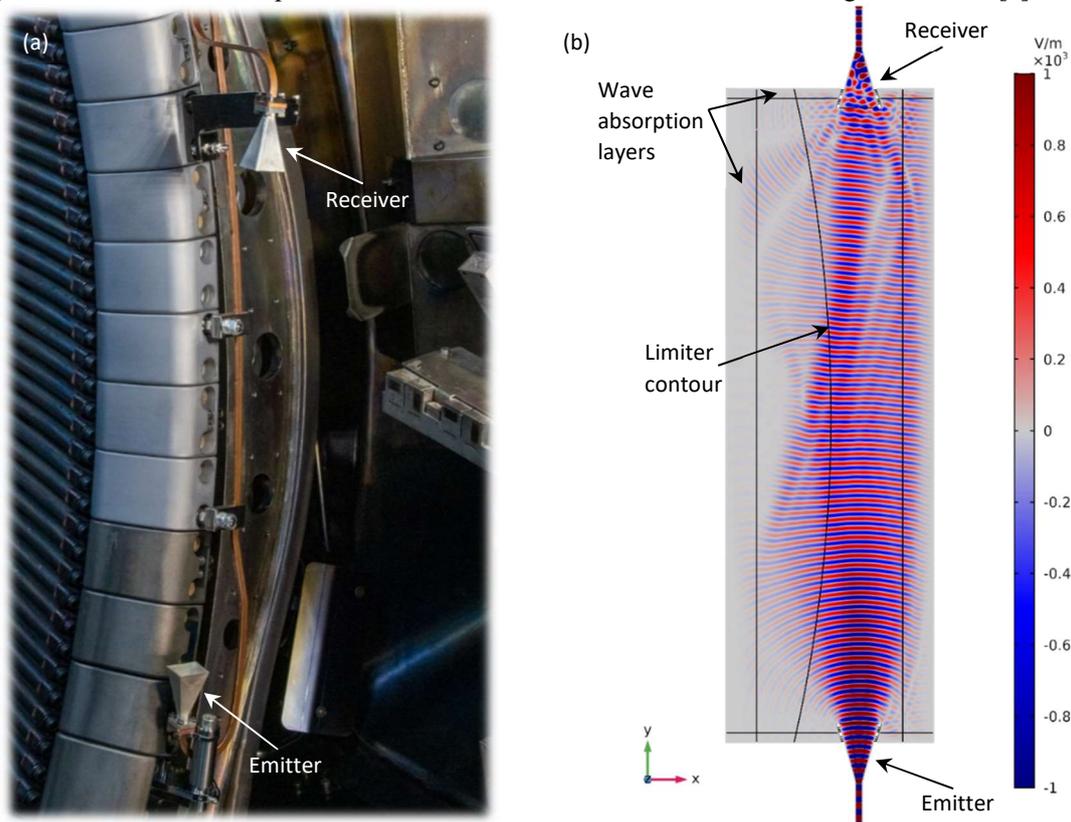

Fig. 1. (a) MILS experimental setup, (b) MILS synthetic diagnostic setup

**2. Model verification**

In previous work, we have verified the correct functioning of the model, i.e. sufficient resolution, solver correctness and error-free parameters definition, by comparing the numerical results to theoretical predictions for some simple cases of plasma density [3]. Here we also perform a simple check in order to ensure that the density perturbations with large gradients over small distances can be sufficiently resolved.

For this test, the density inside the plasma volume is defined as a sum of two contributions. To a constant density background of $n_{bg} = 0.5*10^{18}$ m$^{-3}$, in each model a perturbation of density along y-direction is added. The density along x and z does not vary. The perturbation has a shape of a narrow Gaussian peak in the middle of MILS axis. The peak density $n_{flm}$ and the size (full width at half maximum – FWHM) of the perturbation are varied. For this test and for further studies we choose these parameters in the range within the typical filament characteristics reported from experimental data [5-7].

The phase shift can be calculated theoretically from the formula:

$$\Delta\varphi = \frac{2\pi d}{\lambda_{vac}}\left(1 - \frac{1}{d}\int_0^d N(x)\,dx\right) = \frac{2\pi d}{\lambda_{vac}}\left(1 - \frac{1}{d}\int_0^d \sqrt{1 - \frac{\omega_p^2}{\omega^2}}\,dx\right) = \frac{2\pi d}{\lambda_{vac}}\left(1 - \frac{1}{d}\int_0^d \sqrt{1 - \frac{\left(n_{bg} + n_{flm}*exp\left(-\frac{y^2}{2\sigma^2}\right)\right)}{n_c}}\,dx\right) \quad (1)$$

where $d$ is the MILS axis length, $\lambda_{vac}$ is the wavelength in vacuum, $N$ is the refractive index, $\omega = 2\pi f$ is the angular frequency of the wave, $\omega_p$ is the plasma frequency, $\sigma = FWHM/(2\sqrt{2*\log(2)})$ is the Gaussian function variance.

The integration of eq. (1) is done numerically, since there is no analytical expression for this integral. The theoretical values are compared to the simulation results (Fig. 2a). The mean deviation from the theory is 0.035° and the maximum deviation is 0.22° (at $n_{flm}= 1.5*10^{18}$ m$^{-3}$, $FWHM = 3$ cm), which means that the numerical errors are extremely low.

A second test is done with a 2D Gaussian perturbation, located at $x = 0$ and $y = 0$ m, i.e. in the middle between the MILS antennas. The density along z is constant. The modelling results are compared with theoretical calculation done with the same eq. (1) and a large deviation is observed for nearly all points (Fig. 2b). The discrepancy is caused by the fact that in this case only a part of the detected beam passes through the maximum of the perturbation (along the MILS axis), while some rays only go through the periphery of the filament and experience lower phase shift. The largest effect is observed for the smallest filament size, where this effect is the most significant.

When the phase value is normalized by a ratio of the filament size and a chosen coefficient (chosen for the best match to theory and representing a radial stretch of a filament, sufficient to cause a full phase shift of the whole collected beam), good matching to the theory can be restored (Fig. 2c):

$$\Delta\varphi_{norm} = \Delta\varphi * \frac{3.3\,cm}{FWHM} \quad (2)$$

The successful model verification allows us to use the model for further studies and to estimate the numerical errors.

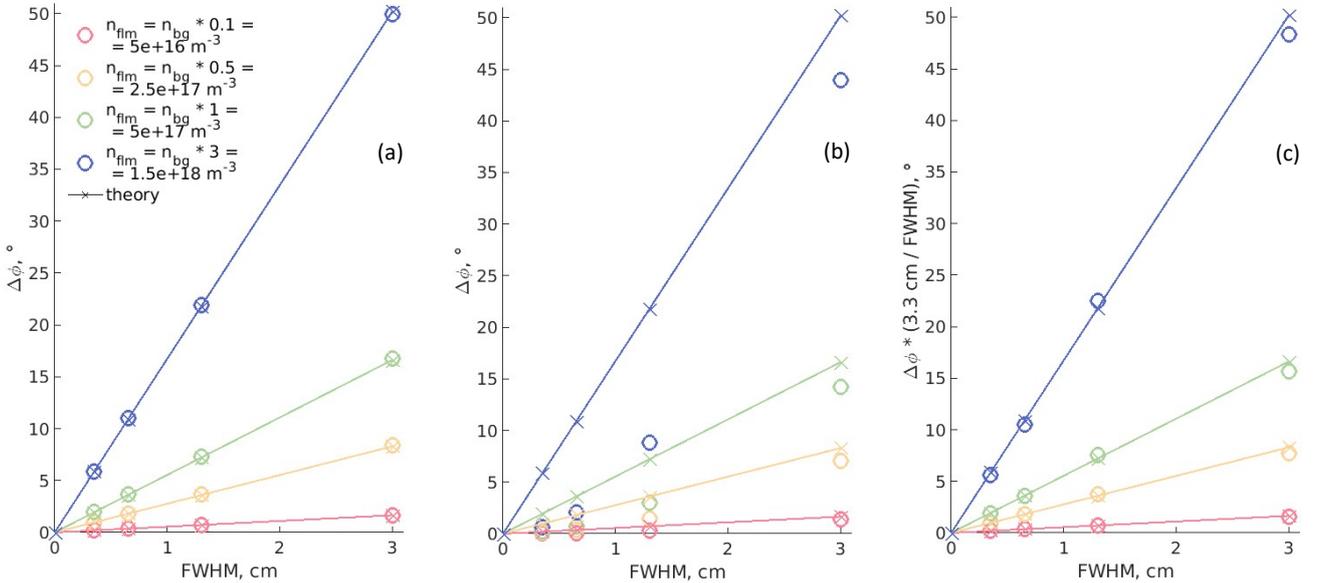

Fig. 2. (a) Numerical and theoretical calculation of phase shift caused by a 1D Gaussian density perturbation along MILS axis. Numerical and theoretical calculation of phase shift caused by a 2D circular filament: (b) phase shift, (c) normalized phase shift, see eq. (2).

## 3. Modelling of filament propagation

In this section we consider a simplified description of a filament-like structure propagating in the SOL. The background density is defined as a monotonic exponentially decreasing density profile same as in [3], described by 3 parameters $(n_{lim}, a_{in}, a_{out})$: $n_{bg} = n_{lim} * exp(a * x)$, where $n_{lim}$ is the density at the limiter and $a$ is the inversed decay length, different for the plasma in front of the limiter $a = a_{in}$ and in the limiter shadow $a = a_{out}$.

A filamentary perturbation is added to the background, with the following simplifying assumptions:
- a single filament is considered
- no poloidal speed, only radial propagation
- no variation of background and filament density along z (toroidally), in 2 cm wide measurement region
- circular or bean-shape x-y cross-section
- 2D Gaussian distribution of filament density
- filament density either constant or exponentially decreasing when radially moved
- far from the separatrix, because in its vicinity an ELM perturabtion of density looks less similar to a filament [6].

While the real filaments can deviate from the adopted assumptions, we want to qualitatively reproduce such type of density perturbation and analyse the sensitivity of MILS to it. Large filaments can be relevant to experimental conditions during a type-I ELM, while smaller ones would be closer to small ELMs or inter-ELM filaments.

*3.1 Filament with radially constant density*

In this test, a filament is moved radially and its density stays constant at all radial positions. It leads to the filamentary distortion being insignificant at the locations with high background density. Therefore the radial range of the filament detection can differ depending on the background density profile and on the density of the filament itself. A scan of the filament density and size, as well as of the background density profile, is performed, in order to find systematic dependencies.

In Fig. 3 the phase-power diagram of MILS synthetic diagnostic output is shown. The database, constructed with monotonic density profiles (described in [3]), is shown with points in gray. Several points of the database serve as starting points for this study and each trajectory of a filament starts from such a point. Arrows show the filament propagation from negative to positive x, i.e. towards the wall. We consider the filament as "detected", when the phase and power values lie outside of the errorbar region of the initial point. For each filament the detection region is given on the plot.

Four examples are given (in blue) of a filament with the same characteristics, which propagates through plasma with different background density. Their trajectories in the phase-power space have slightly similar shapes and all exhibit quite large deviation from their initial points. When the three examples are considered, which start from the same point (solid lines), it can be observed that their trajectories repeat exactly the same shape, while the size of the trajectory scales with the filament density.

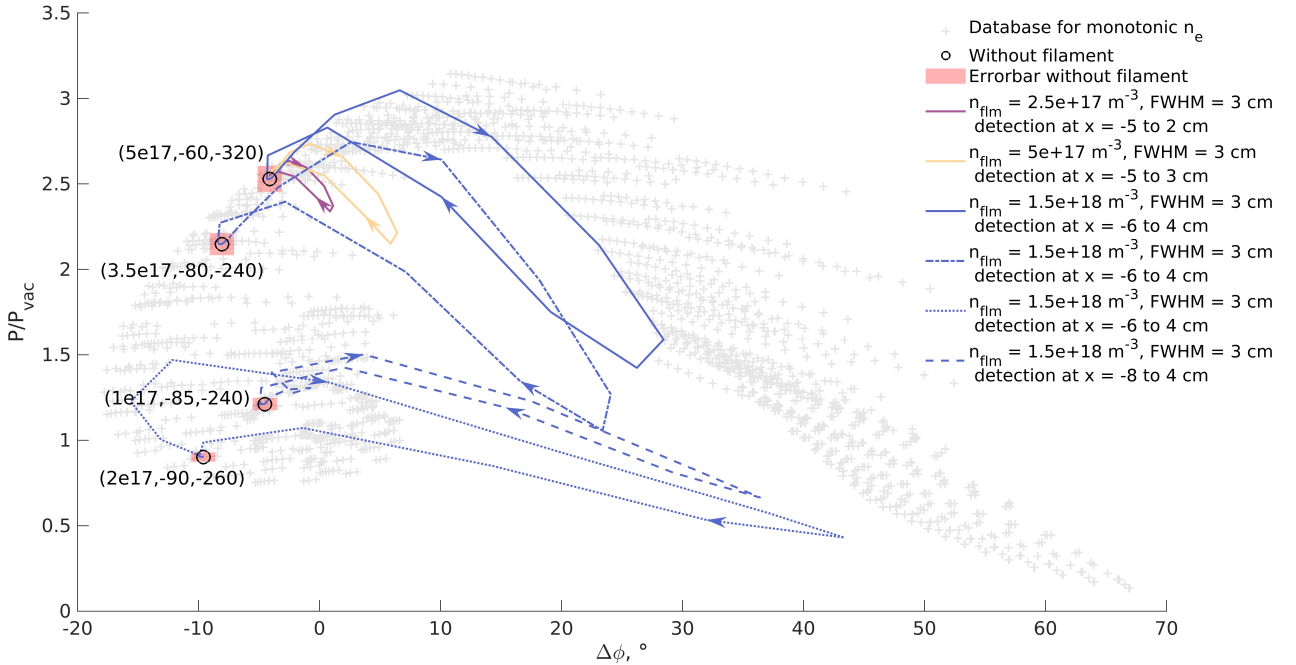

Fig. 3. Phase – power diagram with points from monotonic density database plotted in the background, some of which are the starting points of tests with a filament, labeled with their density profile in the format $(n_{lim}, a_{in}, a_{out})$. Influence of a filament presence is shown as a trajectory, arrows show the filament movement from dense plasma to the edge. Detection range along x is given for each filament (x = 0 cm is the MILS axis). This figure is for filaments with constant density at each radial position.

It is also possible to establish quantitative correlations between the filament properties and the MILS signal variation. In Fig. 4 the maximum deviations of $\Delta\varphi$ and $P/P_{vac}$ from the initial point are plotted as functions of the filament integral density $n_{intg} = FWHM * n_{flm}$. There is a clear correlation (linear in logarithmic scale) for both phase and power, with the approximation functions $\log_{10} \Delta\varphi_{dev} = 1.434 * \left(\log_{10} \frac{n_{intg}}{10^{18}} + 0.41\right)$ and $\log_{10} P/P_{vac\,dev} = 1.003 * \left(\log_{10} \frac{n_{intg}}{10^{18}} - 0.82\right)$, respectively. From the same plot, the limits of the filament detection can be derived, in terms of integral density. For a given lower limit on the detectable phase and power deviation ($\Delta\varphi = 1°$ and $P/P_{vac} = 0.03$ are taken as an example), the minimum detectable filament integral density can be obtained ($\approx 3*10^{17}$ cm*m$^{-3}$ in this example).

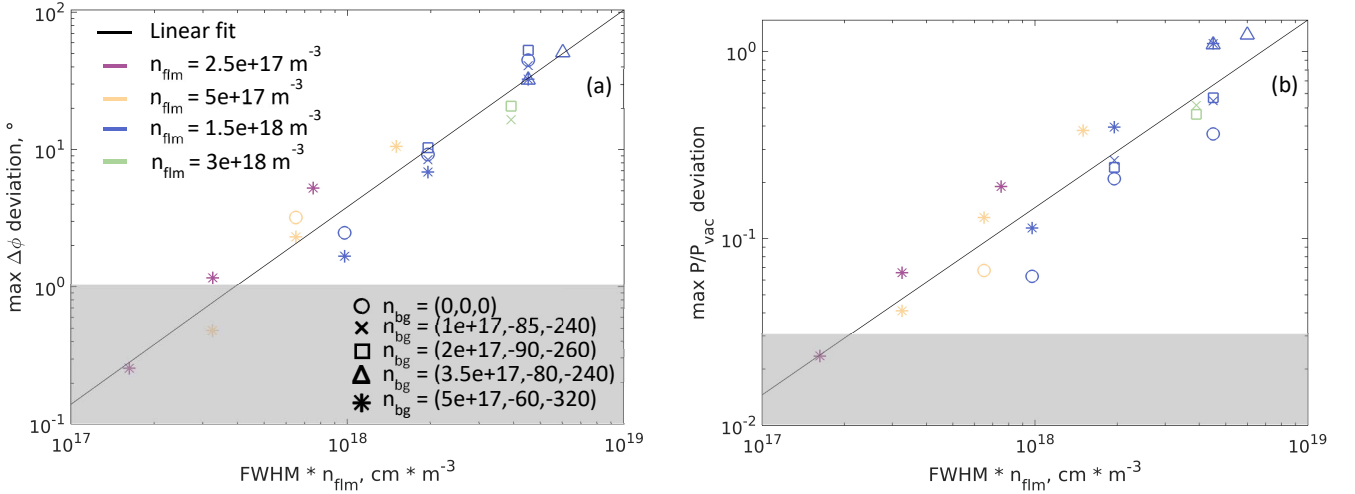

Fig. 4. Correlation of the (a) phase and (b) power maximum deviation with the filament integral density. The shaded area illustrates an example of the border of the filament detection.

*3.2 Filament with radially decaying density*

Another study has been done, where the same circularly shaped filament is tracked during the same radial movement, but the density of the filament changes at each radial step. Its density is proportional to the background density at each point.

Results of the simulations are shown in Fig. 5 in the same way as in the previous subsection. Large qualitative differences can be seen in the shapes of the trajectories compared to Fig. 3. It means that on the qualitative level we are able to distinguish the two used theoretical descriptions of the filament propagation.

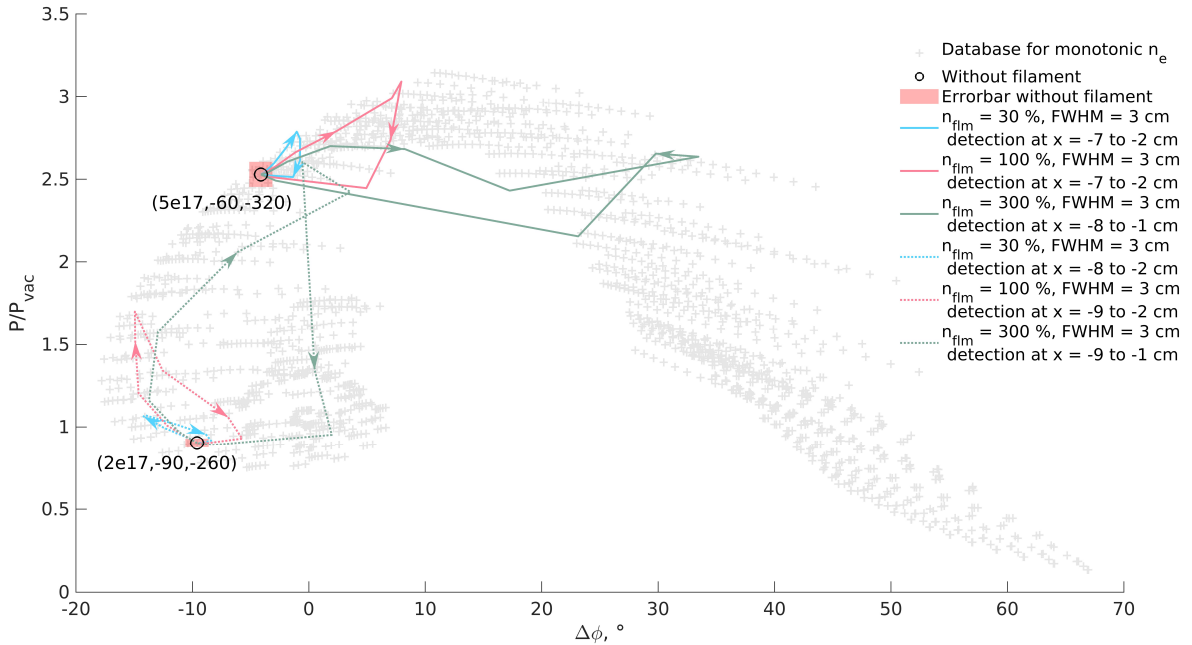

Fig. 5. Examples of filaments with density decaying with radial position (phase – power diagram as in Fig. 3).

The same (linear in the logarithmic scale) correlation is observed between the maximum deviation of the phase and power and the integral density of the filament, with the approximation functions $\log_{10} \Delta\varphi_{dev} = 0.675 * \left(\log_{10} \frac{n_{intg}}{10^{18}} + 0.52\right)$ and $\log_{10} P/P_{vac\,dev} = 1.186 * \left(\log_{10} \frac{n_{intg}}{10^{18}} - 1.41\right)$, respectively. Different to the previous case, the typical radial location of the filament, when it introduces the largest perturbation to MILS measurements, is not in the limiter shadow. For a filament of constant density, the maximum is almost always at x = 0 cm, i.e. at the MILS axis. It is explained by the fact that such filament becomes prominent only when it reaches the radial position with quite low background density and it affects the MILS signal the most when located at MILS axis. For a filament with radially decreasing density, the location of maximum perturbation is shifted towards the core plasma, with the mean value at x = -4.5 cm, i.e. at 2.5 cm in front of the limiter. Such filament, when moved further to low-density plasma, becomes less dense itself and causes weaker change in the signal of MILS.

*3.3 Circular filament and filament elongated poloidally*

We compare the impact of a circular filament and of a filament with a cross-section shape elongated along the flux surfaces (poloidally), shown in Fig. 6. The density destribution of the elongated filament is defined by the function:

$$n_{flm}(x,y) = n_{max} * exp\left(-\frac{(R-R_0)^2}{2\sigma_r^2}\right) * exp\left(-\left(asin\frac{y}{R_0} - asin\frac{y_0}{R_0}\right)^2 * \frac{R_0^2}{2(\varepsilon*\sigma_r)^2}\right) \qquad (3)$$

where the coordinates x,y are defined differently from their previous usage, they have zero at the center of the concentric circles of constant density, approximating the flux surfaces; $R = \sqrt{x^2 + y^2}$, $R_0 = \sqrt{x_0^2 + y_0^2}$, $x_0$ and $y_0$ are the coordinates of the filament center, $\varepsilon$ is the elongation and $\sigma_r$ relates to the radial size of the filament.

For relatively low (experimentally relevant – see [7]) values of the ratio between the radial and the poloidal size, the trajectory of an elongated filament repeats nearly exactly the trajectory of a circular filament with the same integral density along $y$ (Fig. 6). It happens because the direction of MILS beam propagation is not very different from the directional of poloidal elongation and of the y-direction, because both the refracted beam and the flux surfaces have moderate curvature. With elongation ratio $\gg 1$, it would not be the same, because the beam would not pass throught the filament center along its whole extension.

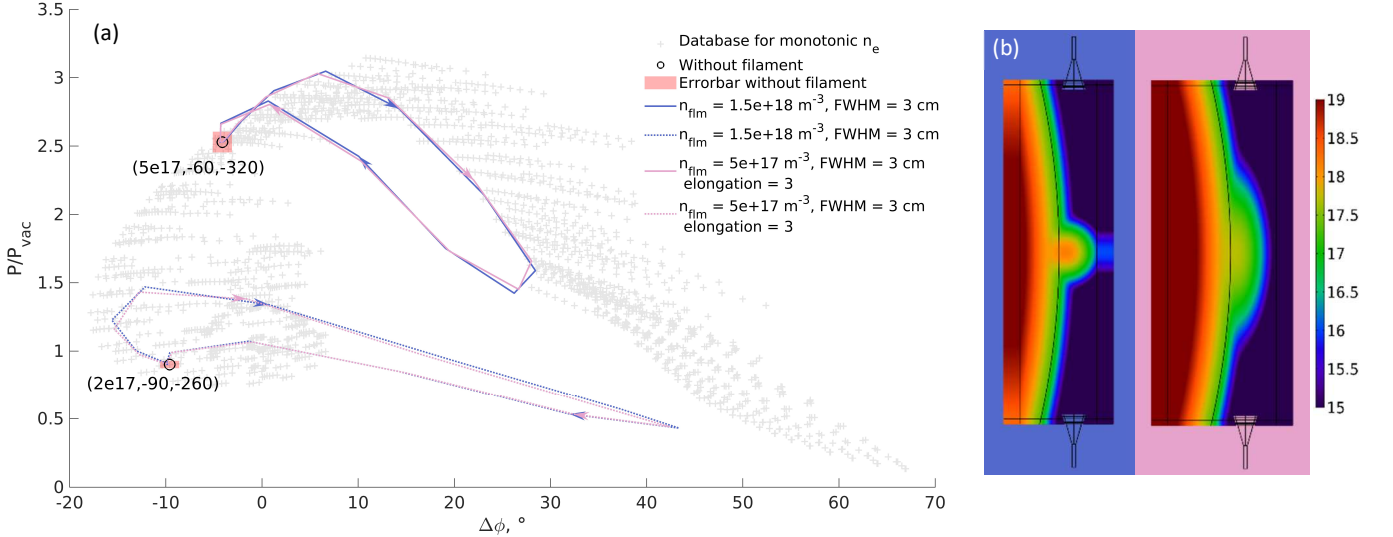

Fig. 6. (a) Comparison of circular and poloidally elongated filaments (phase – power diagram as in Fig. 3). (b) 2D density profiles with circular and poloidally elongated filament shape, with the plot background color corresponding to the trajectory color.

**4. Sensitivity of filament detection to its location**

The tests in Section 3 have been done with a filament moving radially in the middle between the MILS antennas. It is useful to explore the spatial limit of a filament detection in a larger range of its location. Fig. 7a,b show the results of an analysis for one case of the background density (0.5e18,-60,-320), filament with radially decaying density: $n_{flm} = 100\%$ of the background density at each radial position, and with the size $FWHM = 3$ cm. When other values are taken for the background plasma and filament properties, the detection area would be different but qualitatively similar. The results demonstrate that there is a strong reduction of the signal perturbation by a filament at locations further away from $y = 0$ cm. Empty circles show the points where the values are below the chosen detection limit, 1° and 3% for phase and power respectively. If the experimental conditions (noise, plasma turbulence, etc) set this limit higher, the detection area would shrink noticeably in $y$ direction.

The shape of the trajectory of a filamentary density perturbation in the phase-power diagram changes when the radial propagation at $y$ different from 0 is considered (Fig. 7c). However, it is still quite close and it does not change strongly enough to not be distinguished from other shapes, e.g. from the trajectories in Fig. 3.

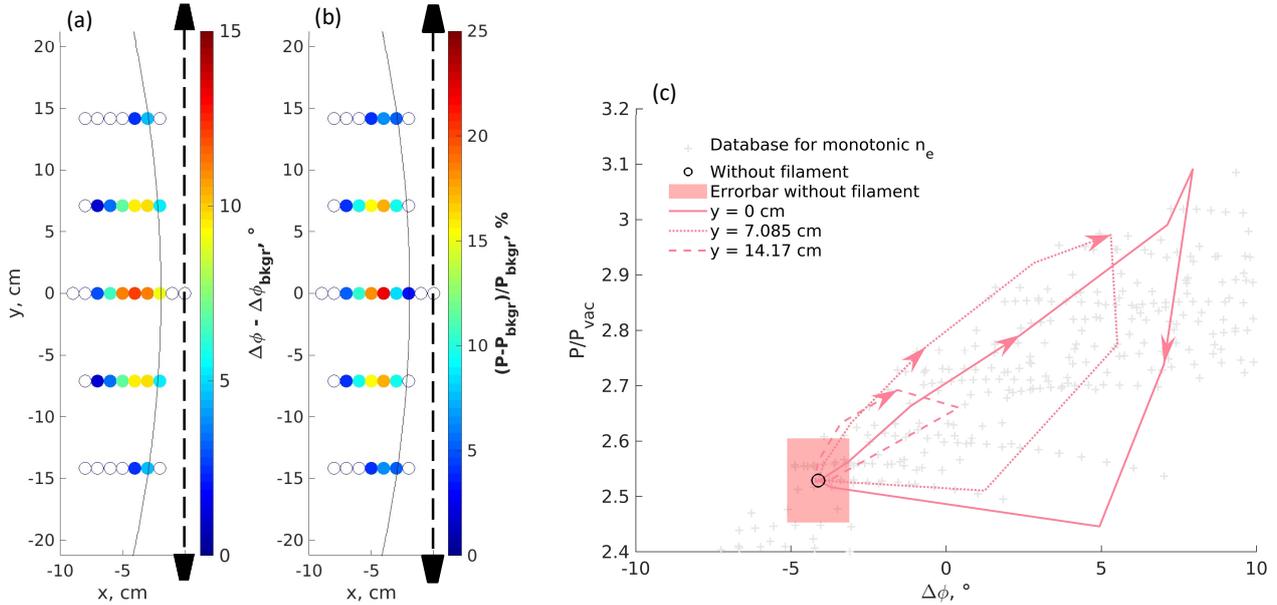

Fig. 7. Example of spatial boundaries of a filament detection by a change in (a) phase and (b) power, where empty circles have values below the detection limit of 1° and 3%, correspondingly. MILS antennas and axis and the limiter contour are schematically shown. (c) Trajectories of a filament propagating radially at different $y$ values (identical at negative $y$, therefore not plotted).

## 5. Experimental examples of ELMs

In experiments on ASDEX Upgrade, the data acquisition rate of 200 kHz allows resolving the temporal evolution of the signal perturbation, caused both by ELMs and by smaller filaments. Examples of such perturbations for the case of ELM filaments, presented in the same way as the modelling results – as trajectories in the phase-power diagram, are shown in Fig. 8a. Many curves resemble the trajectories in Fig. 3, while some look as a case in between the shapes from Fig. 3 and Fig. 5. These qualitative comparisons indicate which theoretical description of the filament density variation during the radial propagation shows results closer to the experimental measurements of MILS. The exponential decay of the filament density with the same decay length as of the background density, used for the results in Fig. 5, seems to be an overestimation. The model with filament density staying constant or an option in between the two considered models, with slower filament density decay, agree better with the considered experimental data for ELMs.

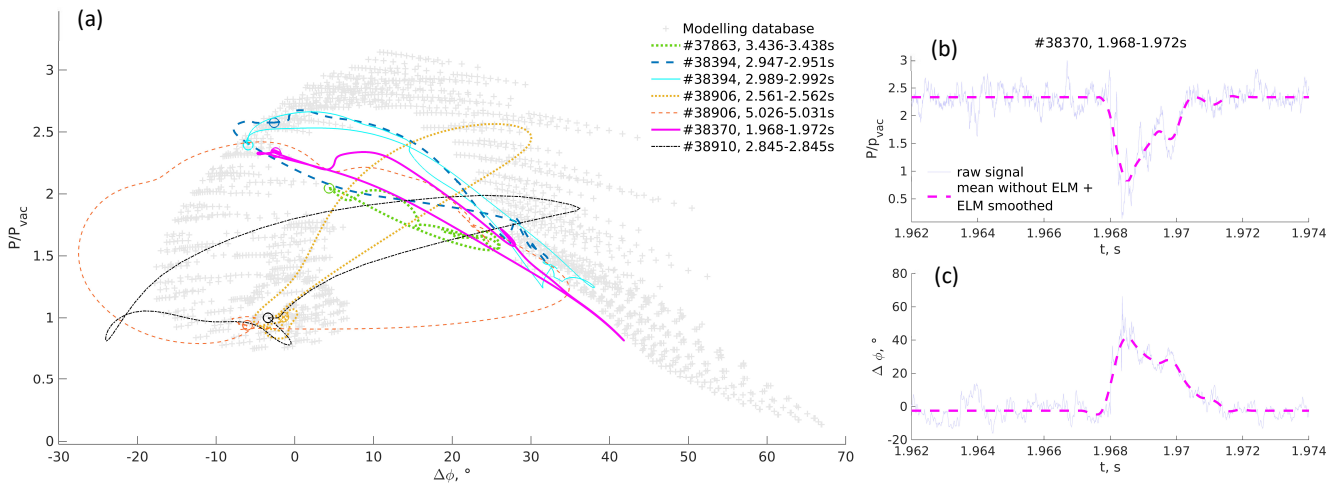

Fig. 8. (a) Examples of experimental MILS signals during ELM filaments in ASDEX Upgrade. Time evolution of (b) power and (c) phase signals for one of the examples in (a), with inter-ELM periods before and after the ELM.

From Fig. 8b,c it can be seen that the level of density perturbation by the turbulence and small filaments is low enough for the meaningful calculation of the backgound mean values and the ELM signals. This example is taken as representative for the typical level of such perturbations, but examples with much lower or much higher perturbations also exist in the MILS data.

From the performed modelling it can be estimated that the filament is detected along ~10 cm radial distance. When this value is divided by the duration of the MILS signal perturbation in experimental data, ~1-2 ms (for the examples of smaller ELMs, corresponding approximately to the simulated filaments), the filament radial speed can be derived, resulting in ~50-100 m/s. While a lot of literature reports higher ELM radial velocities [7], such results are moslty given for the velocity at the separatrix and in the near SOL. In the far SOL, the ELM filament velocity was observed to decrease [8].

More advanced analysis is required to make improved comparisons between the experiment and the theories. The presented study does not cover the filaments, which have non-negligible poloidal speed. Their analysis will have an influence both on the interpretation of the shape of the filament trajectory in the phase-power diagram with respect to theories describing its properties and on the evaluation of the filament velocity. The current study is limited to the qualitative investigation of MILS sensitivity to filaments, and the refined comparison to the experiment is foreseen in the future work.

## 5. Conclusions

Filamentary perturbations of density in the far SOL are typical for experimental tokamak conditions. MILS, as a diagnostic aimed at density meaurements in that region, is capable of detecting the density oscillations with high temporal resolution. For the interpretation of the observed oscillations it is essential to study how a filamentary perturbation affects the MILS signals. In this paper that question is examined by using basic theoretical descriptions of a single filament in the far SOL.

The performed analysis helps to outline the boundaries of the filament detection by MILS for the filament size, density and poloidal-radial location. Qualitative features like the trajectory of the filamentary perturbation signal in the phase-power diagram can be assosiated with some characteristic assumptions used in the theoretical definition of the simulated density perturbation. Quantitative values can be obtained from the derived scalings of the measured quantities, phase and power, with the integral density of a filament. The velocity of the filament (only filaments with dominant radial velocity component are considered in this study) is indicated by the duration of the MILS signal perturbation, with the typical travelled length in the field of view of the diagnostic estimated from modelling.

This work also displays the sensitivity of MILS to the density perturbation in far SOL in general, since a single filament and a turbulent structure distributed along the whole MILS path can have similar influence on the measured signal. The integral value of the density perturbation can be estimated by the level of MILS signal oscillation from the found scalings.

MILS synthetic diagnostic can utilize any 3D density distribution as an input, for example from another modelling or from the experiment. With such forward modelling it can be possible to study numerically and validate experimentally more advanced theories of density perturbation and turbulence in far SOL.

**Acknowledgements**

This work has been carried out within the framework of the French Federation for Magnetic Fusion Studies (FR-FCM) and EUROfusion Consortium and has received funding from the Euratom research and training programme 2014–2018 and 2019–2020 under Grant agreement No. 633053. The views and opinions expressed herein do not necessarily reflect those of the European Commission.